\documentclass[sn-mathphys,Numbered]{sn-jnl}

\usepackage{graphicx}%
\usepackage{multirow}%
\usepackage{amsmath,amssymb,amsfonts}%
\usepackage{amsthm}%
\usepackage{mathrsfs}%
\usepackage[title]{appendix}%
\usepackage{xcolor}%
\usepackage{textcomp}%
\usepackage{manyfoot}%
\usepackage{booktabs}%
\usepackage{algorithm}%
\usepackage{algorithmicx}%
\usepackage{algpseudocode}%
\usepackage{listings}%

\theoremstyle{thmstyleone}%
\theoremstyle{thmstyletwo}%

\theoremstyle{thmstylethree}%

\graphicspath{{figures/}{./}}
\usepackage{doi} 
\usepackage{orcidlink}
\usepackage{xspace}                    
\usepackage{bbm} 
\usepackage[numbers,sort&compress]{natbib}
\usepackage{xparse}
\NewDocumentCommand{\arxiv} %
{r [: u{ [} u{]]} }{[\href{http://arxiv.org/abs/#2}{arXiv:#2}~[#3]]}
\NewDocumentCommand{\arxivold} {r[]}{[\href{http://arxiv.org/abs/#1}{#1}]}
\NewDocumentCommand{\arXiv} %
{r [: u{ [} u{]]} }{[\href{http://arxiv.org/abs/#2}{arXiv:#2}~[#3]]}
\NewDocumentCommand{\arXivold} {r[]}{[\href{http://arxiv.org/abs/#1}{#1}]}
\newcommand{\eg}{\textit{e.g.}\xspace}
\newcommand{\ie}{\textit{i.e.}\xspace}
\newcommand{\etal}{\textit{et al.}\xspace}

\newcommand{\dis}{\displaystyle}

\newcommand{\calK}{\mathcal{K}}
\newcommand{\calR}{\mathcal{R}}

\newcommand{\e}{\mathrm{e}}
\newcommand{\ii}{\mathrm{i}}

\newcommand{\jth}{{(j)}}
\newcommand{\LambdaEfimov}{\ensuremath{\Lambda_\ast}}

\newcommand{\absr}{|r_0|}
\newcommand{\absrm}{|r_0^{-1}|}
\newcommand{\boxing}[2]{\rule[-0.75ex]{0ex}{3.5ex}#2}
\newcommand{\xithr}{\xi_{\mathrm{thr}}}
\newcommand{\xiZB}{\xi_{\mathrm{ZB}}}
\newcommand{\kappatwoZB}[1][]{\kappa_{2\mathrm{ZB}}^{-\:#1}}
\newcommand{\kappathr}{\kappa_{\mathrm{thr}}}
\newcommand{\kappaQU}{\kappa_{3\mathrm{QU}}}

\newcommand{\multi}{\!\cdot\!}

\newcommand{\B}{\mathrm{B}}
\newcommand{\N}{\mathrm{N}}
\newcommand{\twoB}{\ensuremath{2\B}\xspace}
\newcommand{\threeB}{\ensuremath{3\B}\xspace}
\begin{document}

\title{Universality for Three Bosons with Large,
  Negative Effective Range: Aspects and Addenda}


\author*[1]{\fnm{Harald W.} \sur{Grie\3hammer}\orcidlink{0000-0002-9953-6512}}\email{hgrie@gwu.edu}



\affil*[1]{\orgdiv{Institute for Nuclear Studies, Department of Physics},
  \orgname{The George Washington University}, \orgaddress{
    \city{Washington DC}, \postcode{20052}, 
    \country{USA}}}




\abstract{Resummed-Range Effective Field Theory is the consistent
  non-relativistic Effective Field Theory of point interactions in systems
  with large two-body scattering length $a$ and an effective range $r_0$ large
  in magnitude but negative. Its leading order is non-perturbative, and its
  observables depend only on the dimensionless ratio $\xi:=2r_0/a$ once
  $|r_0|$ is chosen as base unit. This presentation highlights aspects for
  three identical spinless bosons and adds details to a previous
  discussion~\cite{Griesshammer:2023scn}. At leading order, no three-body
  interaction is needed. A ground state exists only in the range
  $0.366\ldots\ge\xi\ge-8.72\ldots$, and excited states display
  self-similarity and Discrete Scale Invariance, with small corrections for
  nonzero $r_0$.}


\maketitle
\section{Introduction}

Two-body systems with shallow poles and effective ranges which are large in
magnitude and negative occur most likely when $k\cot\delta$ has a large and
negative curvature, typically induced by a resonance
just above zero momentum, like in narrow Feshbach resonances, the
$D^*_s(2317)$ and some $\Xi\N$ and $\Xi\Xi$ systems~\cite{vandeKraats:2022kde,
  PhysRevLett.93.143201, Matuschek:2020gqe, Gasparyan:2011kg,
  Haidenbauer:2015zqb, Haidenbauer:2021zvr}. Habashi and
collaborators~\cite{Habashi:2020qgw,Habashi:2020ofb,vanKolck:2022lqz} studied
such two-body systems by formulating a consistent non-relativistic Effective
Field Theory (EFT) of one momentum-independent and one momentum-dependent
contact interaction between two identical spinless bosons. This presentation
discusses some aspects of the cornucopia of universal scaling properties in
its \emph{three}-boson sector, adding material to a comprehensive
study~\cite{Griesshammer:2023scn}, which also contains a fuller list of
references.

\section{A Digest of the Formalism}

\begin{figure}[!h]%
\centering
  \includegraphics[width=0.85\linewidth]
  {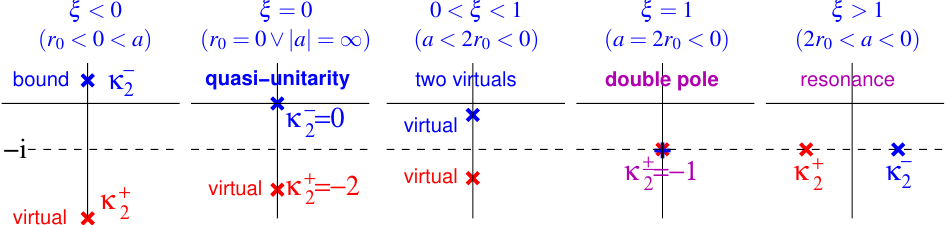}
  
  \caption{(Colour on-line) \twoB poles in the complex momentum plane as $\xi$
    ($\kappa_2^-$) increases (decreases).}
\label{fig:2Bpoles}
\end{figure}

Consider a system of two identical spinless, non-relativistic bosons which are
characterised by a scattering length $a$ and effective range $r_0$. Once
$\absr$ is chosen as base unit for distances and momenta, its observables are
universal in the sense that they depend only on the dimensionless ratio
\begin{equation}
  \xi:=\frac{2r_0}{a}\;\;.
\end{equation}
In the usual version called ``Short-Range EFT'', one considers $|r_0|\ll|a|$
($\xi\to0$) as perturbation and needs a stabilising three-boson (\threeB)
interaction at leading order, leading to Efimov's Discrete Scale
Invariance~\cite{Efimov:1970zz, Efimov:1971zz, Efimov:1973awb, Efimov:1978pk};
see \eg ref.~\cite{Hammer:2019poc} for a recent review. Here, we consider
``Resummed-Range Effective Field Theory'', the extension of Short-Range EFT to
include large, and hence non-perturbative, $\xi$. In the $2\B$ sector, its
leading order is of course still non-perturbative, and the two interactions
are tuned to reproduce both a large scattering length $a$ and an effective
range $r_0<0$ large in magnitude but
negative~\cite{Habashi:2020qgw,Habashi:2020ofb,vanKolck:2022lqz}. The theory
is also renormalisable provided the effective range is negative so that the
Wigner bound is satisfied~\cite{Phillips:1997xu,Beane:1997pk,
  Fewster:1994sd,Phillips:1996ae}:
\begin{equation}
    r_0<0\;\;.
\end{equation}
Its leading-order \twoB $S$-matrix takes the ``second-simplest'' non-trivial
form,
\begin{equation}
  S=\frac{K+\ii\kappa_2^-}{K-\ii\kappa_2^-}\;
  \frac{K+\ii\kappa_2^+}{K-\ii\kappa_2^+}\;\;,
\end{equation}
characterised by exactly two poles with positions $\ii\kappa_2^\pm$ in the
complex momentum plane:
\begin{equation}
  \kappa_2^\mp:= - \left[1\mp\sqrt{1-\xi}\right]
  \xrightarrow[\text{un-scale}]{r_0\ll a\;(\xi\to0)}\left(\frac{1}{a},-\frac{2}{\absr}\right)\;\;.
\end{equation}
[This equation includes the Short-Range EFT limit in un-rescaled variables, in
which one pole becomes shallow, the other one deep.] Their evolution from one bound and one virtual state
at $\xi<0$ to a resonance for $\xi>1$ is shown in fig.~\ref{fig:2Bpoles}.

\begin{figure}[!t]%
\centering
  \includegraphics[width=0.6\linewidth]
  {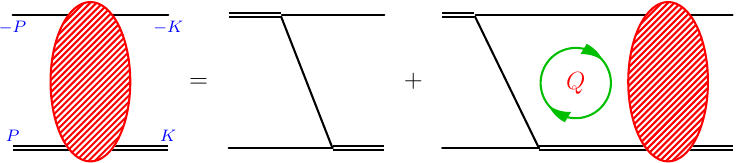}
  
  \caption{(Colour on-line) Diagram of the $3\B$ Faddeev integral
    equation. Double line: $2\B$ amplitude.}
\label{fig:3Binteq}
\end{figure}

In the $3\B$ system, the leading order is of course still nonperturbative. The
corresponding Faddeev equation is shown
in fig.~\ref{fig:3Binteq}. The $j$th bound state at momentum
$\ii\kappa_3^\jth$ is the nontrivial solution to its homogeneous part with
$S$-wave kernel
\begin{equation}
  \label{eq:kernel}
  \calK(-(\kappa^\jth_3)^2;P,Q):= \frac{Q}{P}\;\frac{\dis\ln
  \frac{P^2+Q^2+PQ+(\kappa^\jth_3)^2}{P^2+Q^2-PQ+(\kappa^\jth_3)^2}}
  {\dis\xi+\frac{3Q^2+4(\kappa^\jth_3)^2}{4}+\sqrt{3Q^2+4(\kappa^\jth_3)^2}}\;\;.
\end{equation}
As a practical matter, one discretises the kernel and scans in the (rescaled)
``binding energy'' $(\kappa^\jth_3)^2>0$ for solutions of
\begin{equation}
  \label{eq:det}
  \det[1-\calK(-(\kappa^\jth_3)^2;P,Q)]=0 \;\;,
\end{equation}
on a mesh which weighs low momenta more than higher ones. For example,
\begin{equation}
  P,Q=  (\e^{z_{P,Q}}-1)\;\kappa_2^-(\xi-\epsilon)
\end{equation}
with $\epsilon\approx \pm10^{-[5\dots7]}$ renders robust results of high
accuracy quickly on grids of between $500$ and $2000$ points over a wide range
of ``hard'' cutoffs $\Lambda\in[0.5;3000]\absrm$.

\section{A Curated Selection of Results}

\begin{figure}[!t]%
\centering
  \includegraphics[width=\linewidth]
  {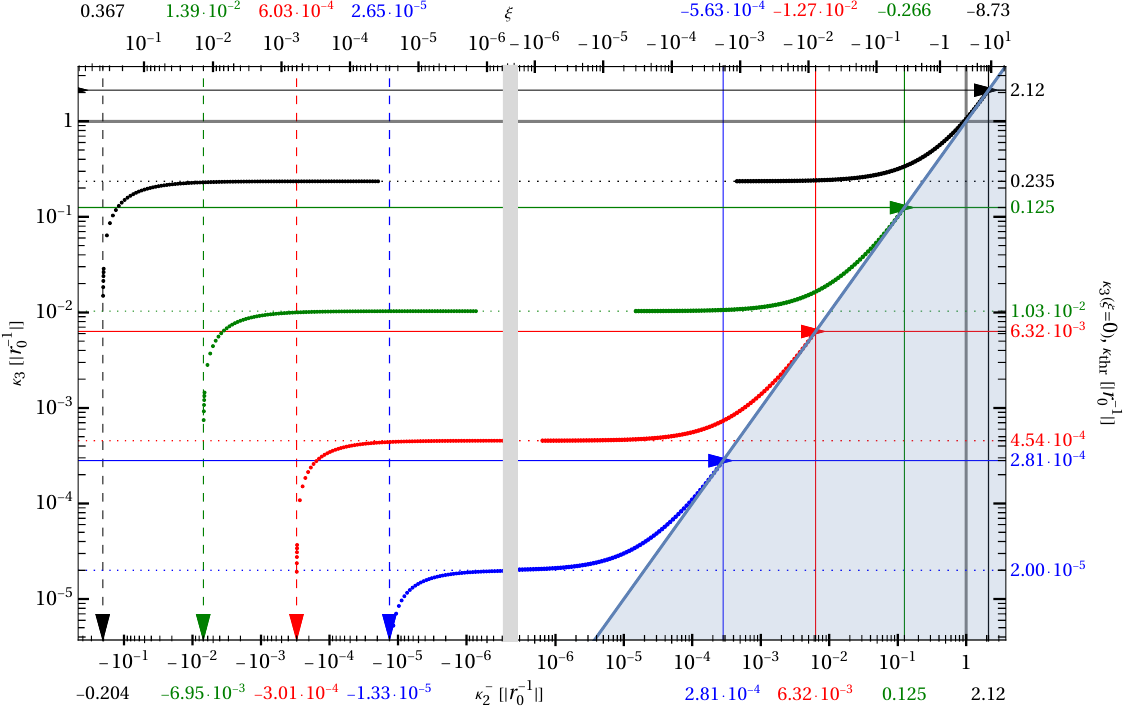}
  
  \caption{(Colour on-line) Binding momenta $\kappa_3$ of the \threeB system
    in $|r_0^{-1}|$ as function of the \twoB binding momentum $\kappa_2^-$
    (bottom scale, in $|r_0^{-1}|$) and of $\xi$ (top scale), for the ground
    state (black) and the $1$st (green), $2$nd (red) and $3$rd (blue)
    excitations. Logarithmic scales on both axes, with positive and negative
    values in $\kappa_2^-$ ($\xi$) stitched together by a gray line. Gray
    region: \threeB state unstable, $\kappa_3<\kappa_2^-$. Right axis:
    $\kappa_3^\jth$ at quasi-unitarity (dotted horizontal lines) and at
    threshold (solid horizontal and vertical lines). Top axis: $\xithr$ at
    threshold and at zero-binding (dashed vertical lines). Bottom axis:
    corresponding $\kappathr$ and $\kappatwoZB$. Arrows point to zero binding
    (vertical), threshold (horizontal), and quasi-unitarity ($\kappaQU$,
    horizontal). All characteristic values are also found in
    table~\ref{tab:table}. The trajectory of the $3$rd excitation is within
    line thickness indistinguishable from the one in Short-Range EFT with the
    same \threeB binding momentum at quasi-unitarity.}
\label{fig:allbindings}
\end{figure}

Figure~\ref{fig:allbindings} shows the central result: the binding momenta
$\ii\kappa_3$ for the ground and first three excited states of the $3\B$
system display what looks by eye to be self-similarity and Discrete Scale
Invariance similar to that found in the Efimov effect of Short-Range
EFT. However, upon closer inspection, small difference surface. The ground
state's binding momentum at threshold is with
$\kappathr^{(0)}\approx2.12\;\absrm$ about $30\%$ smaller than for an Efimov
trajectory with the same \threeB binding momentum at $\xi=0$, where it is
about $3.31\;\absrm$. The difference is still about $15\%$ for the first
excitation. At zero binding, the ground (first excited) state is about $30\%$
($10\%$) further away than the Efimov case. Overall, trajectories of the
ground state and first few excitations are shorter in $\kappa_2^-$ than those
of Efimov trajectories.  And while the tower of states is infinitely high and
deep in Short-Range EFT, no stably-bound state exists in Resummed-Range EFT
outside the range $0.367\ldots \ge\xi\ge-8.726\ldots$ (namely
$-0.204\ldots\le\kappa_2^-\le2.118\ldots$), and excitations appear only for
smaller $|\xi|$ ($|\kappa_2^-|$): There are no bound states at $\xi=-10$, one
at $\xi=-1$, two at $\xi=-0.1$, three at $\xi=-0.01$, and four at
$\xi=-0.0001$.

These findings are however not un-surprising after a look at the kernel,
eq.~\eqref{eq:kernel}. The loop momentum $Q$ dominates
the $2\B$ propagator for $Q\gg\kappa_3,\xi$ and tames its UV limit so that a $3\B$ interaction is not
needed to stabilise the system against collapse. Since there are no external
scales, no bound states emerge. At intermediate
$\frac{3Q^2}{4}+\kappa_3^2\gg \sqrt{3Q^2+4\kappa_3^2}\gtrsim4$, the
effective-range term $\frac{3Q^2}{4}+\kappa_3^2$ dominates over the unitarity
term $\sqrt{3Q^2+4\kappa_3^2}$, and the integral still converges
quickly. Since the binding momentum sets the scale of the loop momentum as
$Q\lesssim\kappa_3$, bound states can be supported for $\kappa_3\lesssim
2$. Indeed, the ground state's binding momentum $\kappa_3^{(0)}\le2.118\dots$
agrees with this estimate. For such states, effective-range effects are large
since $\absr\sim a$ ($\xi\sim1$).

Each bound state maps out a trajectory as function of $\xi$. At the
\textbf{Threshold Point}
$\kappathr:=\kappa_3(\xithr)\stackrel{!}{=}\kappa_2^-(\xithr)$, a \threeB
state emerges from the $2\B$ continuum. As $0>\xi>\xithr$ decreases in
magnitude, both this \threeB and all \twoB sub-systems are bound, until the
binding momentum of the shallowest \twoB bound state vanishes at
\textbf{Quasi-Unitarity} $\kappa_2^-(\xi=0)=0$, namely infinite \twoB
scattering length. Nonetheless, the \threeB state remains bound there,
$\kappaQU:=\kappa_3(\xi=0)>0$. As $\xi\to0$ (and hence the effective range
$r_0\to0$) becomes perturbative around this point,
$\frac{3Q^2}{4}+\kappa_3^2\ll \sqrt{3Q^2+4\kappa_3^2}$ and the leading
contribution becomes identical to Efimov's Short-Range EFT. This predicts that a
self-similar tower of Efimov states appears in which neighbouring trajectories
obey ``Efimov rescaling'', characterised by the transcendental number
$s_0=1.0062378\dots$:
\begin{equation}
  \label{eq:Efimovratio}
  \lim_{j \to \infty}\left(
  \frac{\kappa_3^{(j)}(\kappa_2^-)}{\kappa_3^{(j+1)}(e^{\pi/s_0}\kappa_2^-)}\;,\;
  \frac{\kappaQU^{(j)}}{\kappaQU^{(j+1)}}\;,\;
  \frac{\kappatwoZB[(j)]}{\kappatwoZB[(j+1)]}\;,\; 
  \frac{\kappathr^{(j)}}{\kappathr^{(j+1)}}\right)= 
  e^{\pi/s_0} = 22.6944\ldots\;\;.
\end{equation}
However, there is no need for a $3\B$ interaction since the binding momentum
$\kappa_2^+(\xi\to0)\to-2\;[\absr]$ of the second, virtual \twoB state
continues to provide a \twoB scale for the \threeB system's binding momentum
$\kappaQU:=\kappa_3(\xi=0)>0$. Since a natural scale persists there, the
system is only called close to ``\emph{quasi}-unitarity''. It is from that
r\'egime that the overall graph inherits a semblance of self-similarity.

As $\xi$ turns positive, one proceeds in the Borromean region of a state's
trajectory, where the \threeB system continues to be bound albeit all $2\B$
subsystems are unbound (\ie~$\kappa_2^\pm<0$ virtual). \threeB binding
continues to decrease until finally, the $3\B$ system itself becomes unbound
at \textbf{Zero Binding} $\kappa_3(\xiZB)=0$ for
$\kappatwoZB:=\kappa_2^-(\xiZB)<0$.

Table~\ref{tab:table} contains the binding momenta of the ground state and
first three excitations at the three characteristic points. All these approach
``Efimov scaling'', eq.~\eqref{eq:Efimovratio}, and the last row indeed infers
universal constants of binding momenta for highly excited states. Numerical
and extrapolation uncertainties are detailed in
ref.~\cite{Griesshammer:2023scn}.

\begin{table}[!h]
  \caption{Characteristics of crucial points on the three-body trajectory: zero
    binding, quasi-unitarity and threshold. Values are given for the ground
    state and first three excitations as marked also in
    fig.~\ref{fig:allbindings}, plus extrapolations for Efimov-like
    states. Uncertainties combine extrapolation and numerics.}
 \label{tab:table}
 \begin{tabular}{|l||l|l|l|}
    \hline
    & \multicolumn{1}{c|}{zero binding} & \multicolumn{1}{c|}{quasi-unitarity} & \multicolumn{1}{c|}{threshold}\\
    state&
           \boxing{$\xiZB:=\xi(\kappa_3=0)$}{$
           \kappa_2^-(\kappa_3=0)\;[\absrm]$}& $
                                                                                       \kappa_3(\kappa_2^-=0)\;[\absrm]$&
                                                                                                                          $
                                                                                                                          \kappa_3\stackrel{!}{=}\kappa_2^-\;[\absrm]$\\\hline\hline
    ground    & \boxing{$3.6689(1) \multi 10^{-1}$}{$-2.04318(6)\multi 10^{-1}$} & $2.35412(3)\multi 10^{-1}$ & $2.11862(2) $ \\\hline
    $1$st exc. & \boxing{$1.3855(1) \multi 10^{-2}$}{$-6.9517(5)\multi 10^{-3}$} & $1.03030(5)\multi 10^{-2}$ & $1.25108(1)\multi 10^{-1}$ \\\hline
    $2$nd exc. & \boxing{$6.0279(2) \multi 10^{-4}$}{$-3.0144(1)\multi 10^{-4}$} & $4.53987(1)\multi 10^{-4}$ & $6.320(1)\multi 10^{-3}$ \\\hline
    $3$rd exc. & \boxing{$2.6538(3) \multi 10^{-5}$}{$-1.3269(2)\multi 10^{-5}$} & $2.00039(5)\multi 10^{-5}$ & $2.810(3)\multi 10^{-4}$ \\\hline\hline
    $j\to\infty$  &
                    \boxing{}{\rule[-0ex]{0ex}{3.5ex}
                    $-0.1551(1)\;\e^{-j\frac{\pi}{s_0}}$} & $0.23381(8)\;\e^{-j\frac{\pi}{s_0}}$ & $3.31(2)\;\e^{-j\frac{\pi}{s_0}}$\\\hline
  \end{tabular}
\end{table}

\begin{figure}[h]%
\centering
  \includegraphics[width=0.5\linewidth]
  {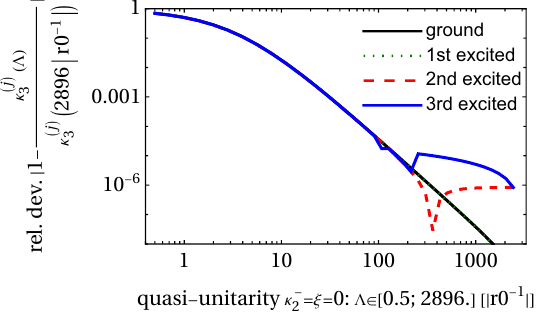}
  
  \caption{(Colour on-line) Dependence of the binding momenta of the ground
    and first three exited states at quasi-unitarity on the cutoff of the
    integral equation eq.~\eqref{eq:det}. For all states, the result at
    $\Lambda\approx10\absr$ differs from the asymptotic one by
    $\lesssim1\%$. At a relative accuracy of $10^{-5}$, numerical
    instabilities enter for the second and third excitation.}
\label{fig:cutoff-dependence}
\end{figure}

Figure~\ref{fig:cutoff-dependence} shows the dependence of the binding momenta
on the cutoff employed to solve eq.~\eqref{eq:det}, confirming that the system
is indeed renormalised (\ie~insensitive to sort-distance Physics)
\emph{without} a stabilising $3\B$ interaction. Notice that the cutoff
variation spans more than two ``Efimov cycles'',
$\Lambda_{\max}/\Lambda_{\min}\approx6000\approx(e^{\pi/s_0})^{2.8}$.
 
\begin{figure}[b]%
\centering
  \includegraphics[height=0.32\linewidth]
  {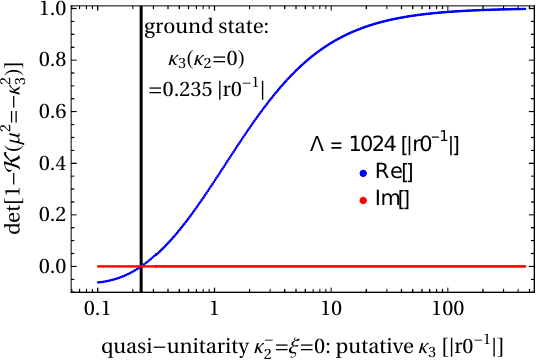}\hspace*{2ex}
  \includegraphics[height=0.32\linewidth]
  {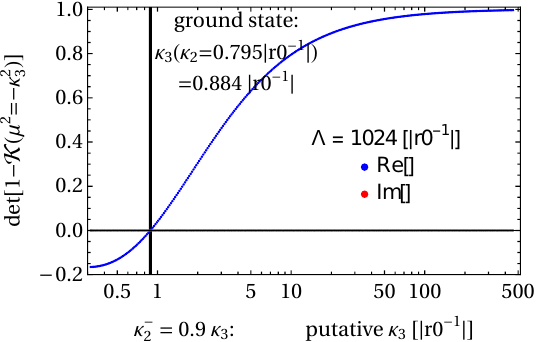}

  \caption{(Colour on-line) Determinant of the bound-state kernel,
    eq.~\eqref{eq:kernel}, at putative values of $\kappa_3$ with fixed
    $\kappa_2^-=0$ (left; quasi-unitarity) and $\kappa_3=\kappa_2^-/0.9$
    (right, parallel to threshold line). The dense $\kappa_3$ distribution
    appears to be a line.  There are no zeroes after
    $\kappa_3^{(0)}(\kappa_2^-=0)=0.235\dots\absr$ (left) and
    $\kappa_3^{(0)}(\kappa_2^-=0.9\kappa_3)=0.884\dots\absr$ (right),
    indicating the absence of more-deeply bound \threeB states.}
\label{fig:search}
\end{figure}

Figure~\ref{fig:search} shows that it is quite unlikely that more-deeply bound
states exist, supporting the argument above. Its two graphs explore zeroes of
the kernel's determinant~\eqref{eq:det}: the left one at quasi-unitarity
$\kappa_2^-=\xi=0$ (vertical gray line in fig.~\ref{fig:allbindings}), the
right one on a parallel to the threshold parametrised by
$\kappa_2^-=0.9\kappa_3$. Both explore more than two ``Efimov cycles'',
$\kappa_{3\max}/\kappa_{3\min}\approx1000\approx(e^{\pi/s_0})^{2.2}$. Both
determinants approach unity for large putative $\kappa_3$, with no indication
of a turn-around. The trained eye will discern that what looks like a line is
actually a dense set of points at which the determinant is sampled.

While scaling is already approximately seen in fig.~\ref{fig:allbindings}, it
becomes more transparent when the trajectory of the $j$th bound
state is normalised to its value at threshold,
\begin{equation}
  \label{eq:kappa2-ratio}
  \calR^\jth\left(x_\kappa:=\frac{\kappa_2^-}{\kappathr^\jth}\right):=
  \frac{\kappa_3^\jth}{\kappathr^\jth}\;\;,
\end{equation}
and expressed as function of the rescaling variable $x_\kappa$ with
$\calR^\jth(x_\kappa=1)\stackrel{!}{=}1$. Inspired by Efimov's universal
function $\Delta(\xi)$~\cite{Efimov:1971zz, Efimov:1973awb, Efimov:1978pk},
one transforms in addition from Cartesian coordinates in the
$(x_\kappa, \calR^\jth(x_\kappa))$ plane to polar ones,
\begin{equation}
    \label{eq:polar}
    \begin{split}
    x_\kappa&=\sqrt{2}\;\rho^\jth(\theta)\;\cos(\theta+\frac{\pi}{4})\;\;,\\
    \calR^\jth(x_\kappa)&=\sqrt{2}\;\rho^\jth(\theta)\;\sin(\theta+\frac{\pi}{4})
    \;\;.
  \end{split}
\end{equation}
The threshold is then normalised to $\rho(\theta=0)=1$, quasi-unitarity is
at $\theta=\frac{\pi}{4}$, and zero binding at $\theta=\frac{3\pi}{4}$.
Using an idea by Gattobigio \etal~\cite{Gattobigio:2019eqw}, the trajectories
are in this form parametrised with residuals of $\ll10^{-3}$ by the $7$
constants of table~\ref{tab:parametrisation} via
\begin{equation}
  \label{eq:parametrisation}
  \rho^\jth_\mathrm{param}(\theta)=
  \exp\left(\sum\limits_{n=1}^7\;c_n^\jth\;\theta^{n/2}\right)\;\;.
\end{equation}
\begin{table}[!h]
  \centering
  \caption{The highly correlated ($|(c_n,c_m)|>0.6$) coefficients of
    $\rho_\mathrm{param}^\jth$ of state $j$ in eq.~\eqref{eq:parametrisation},
    in $\mathrm{rad}^{-n/2}$.  The $68\%$ confidence interval for a prediction
    at a single, new $\theta$ is
    $\rho_\mathrm{param}^\jth(\theta)\pm w_\mathrm{pred}^\jth$.}
    \label{tab:parametrisation}
    \begin{tabular}{|r||c|c|c|c|c|c|c||c|}
    \hline
    state&$c_1^\jth$&$c_2^\jth$&$c_3^\jth$&$c_4^\jth$&$c_5^\jth$&$c_6^\jth$&$c_7^\jth$&$w_\mathrm{pred}^\jth$\\\hline\hline
    $j=0$       &$-4.7545$&$ 4.3710$&$ -7.4404$&$ 11.714$&$ -11.0015$&$ 5.54726$&$-1.12595$   &$0.00009$\\\hline
    $j=1$&$-4.2764$&$ 1.0734$&$ -1.6005$&$ 5.4777$&$ -6.58104$&$ 3.64577$&$-0.768214$  &$0.00003$\\\hline
    $j=2$&$-4.6823$&$ 1.4352$&$ -1.2221$&$ 4.1199$&$ -5.08001$&$ 2.87318$&$-0.613720$  &$0.00003$\\\hline
    $j=3$&$-4.7003$&$ 1.1952$&$ 0.014514$&$ 1.5657$&$-2.44535$&$ 1.53708$&$-0.348975$  &$0.00011$\\\hline
    \end{tabular}
\end{table}
\begin{figure}[!b]%
\centering
  \includegraphics[width=0.53\linewidth]
  {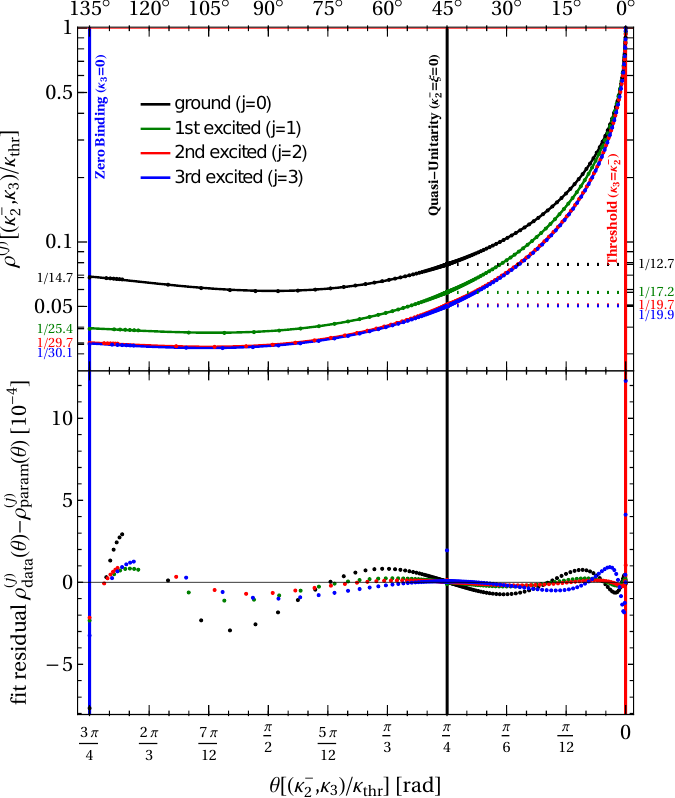}
  
  \caption{(Colour on-line) Top: Data and fit of trajectories of the ground
    state and first three excitations in polar form, eq.~\eqref{eq:polar}, in
    a linear-log plot. The intra-trajectory ratios
    $\kappaQU^\jth/(\sqrt{2}\kappathr^\jth)$ and
    $|\kappatwoZB[\jth]/(\sqrt{2}\kappathr^\jth)|$ are marked on the right and
    left, respectively. To facilitate comparison with
    fig.~\ref{fig:allbindings}, $\theta$ increases from threshold (right) to
    the left. Bottom: Fit residuals on a linear scale, rescaled by $10^4$.}
\label{fig:polar}
\end{figure}

In this parametrisation shown in fig.~\ref{fig:polar}, one clearly sees that a
large effective range (\ie~large $\xi$) has indeed the biggest impact on the
ground state, while its effect on higher excitations decreases. The second and third excitation's trajectories are practically
indistinguishable from each other, and either can be taken to represent an
``Efimov'' trajectory in Short-Range EFT with the same $\kappaQU$.

This similarity also explains that within each trajectory, the ratios between
$\kappa_3^\jth$ at zero binding and threshold, and between quasi-unitarity and
threshold, approach the Efimov-scaling predictions of Short-Range EFT; see
values on the axes of fig.~\ref{fig:polar}.

However, there is an important difference. In Resummed-Range EFT, a \threeB
interaction is needed to stabilise the system. By dimensional transmutation,
this renormalisation leads to a dimensionful scale parameter $\LambdaEfimov$
which fixes the position of the Efimov tower of \threeB states, but whose
value may differ in different systems since this theory has at leading order
no knowledge of $r_0$. In contradistinction, the position of the tower is in
Resummed-Range EFT a parameter-free prediction once the positions of the two
\twoB poles are determined. This different behaviour of the two EFTs is, of
course, a consequence of the fact that Short-Range EFT has no remaining
natural scale once the \twoB binding is zero, while the second pole at
$\kappa_2^+=-2$ still provides one in Resummed-Range EFT. Therefore,
Resummed-Range EFT can be interpreted as an underlying theory to Short-Range
EFT. It \emph{predicts} $\LambdaEfimov$ in units of $\absrm$ for $r_0<0$ when
one matches the \threeB binding momentum of the same state in the two versions
at quasi-unitarity. This value (in regularisation with a ``hard'' cutoff) is
actually near-identical for all states, with a less-than $1\%$ deviation even
for the ground state:
\begin{equation}
  \label{eq:LambdaEfimovDetermined}
  \LambdaEfimov^{(j\ge2)}= 0.610206(1)\;\absrm\;\;.
\end{equation}

\section{Conclusion}

In Resummed-Range EFT, properties of few-body systems are determined by two
parameters: the two-body effective range $r_0$ and the dimensionless ratio
$\xi=2r_0/a$, where $a$ is the \twoB scattering length. As long as $r_0<0$,
this EFT is self-consistent and renormalisable at leading order without a
three-body interaction. While its spectrum appears at first glance quite
self-similar and close to a truncated Efimov spectrum, noticeable
$\xi$-dependence exists for the ground state and lowest excitations. Still,
the spectrum approaches Efimov's exact Discrete Scale Invariance for higher
excitations.

If Short-Range EFT is interpreted as low-energy version of Resummed-Range EFT,
then the absolute position of the Efimov tower is fixed, and the corresponding
Efimov scale is $\LambdaEfimov=0.610206(1)\;\absrm$ (in a renormalisation
scheme with ``hard'' cutoff regularisation); see
eq.~\eqref{eq:LambdaEfimovDetermined}. It will be quite interesting to put
this prediction to the test in systems where Resummed-Range EFT applies.

Further details about these findings and results on scattering a boson on a
bound $2\B$ state in Resummed-Range EFT can be found in
ref.~\cite{Griesshammer:2023scn}.  The fate of the \threeB trajectory in the
sub-threshold and unbound regions will be discussed in future
publications~\cite{future}.  Obvious extensions include systems of
three identical fermions; bosons with different masses; more than three
particles; and higher-order effects, \eg~from a \twoB shape
parameter. Finally, the predictions (with truncation estimates from
higher-order effects) should be confronted with data of suitable systems in
Atomic, Nuclear and Particle Physics which are characterised by a large but
negative effective two-body effective range and a large-in-magnitude
scattering length.

\backmatter

\bmhead{Acknowledgements}

It continues to be a pleasure to collaborate with Ubirajara van Kolck.
I am grateful to the organisers and participants of EuroFewB 2023 in Mainz
for spirited, stimulating and profound discussions, a delightful atmosphere,
and for indulging such a topic as the last plenary presentation.
Instrumental for this research were the warm hospitality and financial support
for stays at IJCLab Orsay and at the Kavli Institute for Theoretical Physics
which is supported in part by the National Science Foundation under Grant
No.~NSF PHY-1748958.
This material is based upon work supported in part by the U.S.~Department of Energy, Office of Science, Office of Nuclear Physics, under award DE-SC0015393.

\subsection*{Author Contributions}

The single author contributed all of the effort to this presentation, and none
more. hg.

\subsection*{Data Availability Statement}

All data underlying this work are available in full upon request from the author.

\subsection*{Competing Interests}

The author declares no competing interests.


\end{document}